\documentclass[fleqn,usenatbib]{mnras}

\usepackage{newtxtext,newtxmath}
\usepackage[T1]{fontenc}

\DeclareRobustCommand{\VAN}[3]{#2}
\let\VANthebibliography\thebibliography
\def\thebibliography{\DeclareRobustCommand{\VAN}[3]{##3}\VANthebibliography}

\usepackage{graphicx}	% Including figure files
\usepackage{amsmath}	% Advanced maths commands
\usepackage[dvipsnames]{xcolor}
\title[Power spectrum in light of JWST observations]{Primordial power spectrum in light of JWST observations of high redshift galaxies}

\author[P. Parashari, R. Laha]{
Priyank Parashari,\thanks{E-mail: ppriyank@iisc.ac.in} Ranjan Laha\thanks{E-mail: ranjanlaha@iisc.ac.in}
\\
Centre for High Energy Physics, Indian Institute of Science,
 C.\,V.\,Raman Avenue, Bengaluru 560012, India
}

\date{Accepted XXX. Received YYY; in original form ZZZ}

\pubyear{2023}

\begin{document}
\label{firstpage}
\pagerange{\pageref{firstpage}--\pageref{lastpage}}
\maketitle

\begin{abstract}
Early data releases of JWST have revealed several high redshift massive galaxy candidates by photometry, and some of them have been confirmed spectroscopically. We study their implications on the primordial power spectrum. In the first part, we use the CEERS photometric survey data, along with respective spectroscopic updates, to compute the cumulative comoving stellar mass density. We find that a very high star formation efficiency (unlikely in various theoretical scenarios) is required to explain these observations within Lambda cold dark matter ($\Lambda$CDM) cosmology. We show that the tension can be eased if the primordial power spectrum has a blue tilt. In the second part, we study spectroscopically confirmed galaxies reported in the JADES survey to investigate their implications on a red-tilted primordial power spectrum. We estimate the star formation efficiency from an earlier observation at similar redshift by {\it Spitzer}, and find that the star formation efficiency is an order of magnitude smaller than required to explain the CEERS photometric observations mentioned earlier. Using the estimated star formation efficiency, we find the strongest constraints on the red tilt of the power spectrum over some scales. Our study shows that JWST will be an excellent probe of the power spectrum and can lead to novel discoveries.
\end{abstract}


%%%%%%%%%%%%%%%%%%%%%%%%%%%%%%%%%%%%%%%%%%%%%%%%%%

%%%%%%%%%%%%%%%%% BODY OF PAPER %%%%%%%%%%%%%%%%%%
\section{Introduction}\label{sec:intro}
Successful operation of the JWST has made it possible to directly observe a large number of galaxies formed very early in the Universe using Early Release Observations program~\citep{2023MNRAS.519.1201A,2022arXiv220711671M,2023ApJ...942L...9Y,2023MNRAS.518.4755A,2023MNRAS.518L..19R}, Early Release Science (ERS) and GLASS-ERS programs ~\citep{Castellano:2022ikm,2022arXiv220801612H,2022ApJ...938L..15C,2022arXiv221010066K}, and  Cosmic Evolution Early Release Science (CEERS) survey~\citep{2022arXiv220712446L,2022ApJ...940L..14N,2022ApJ...940L..55F}.These observations are useful to understand the cosmological structure formation at high redshifts ($z\,\gtrsim 7$) or to discover new physics.

 Recent observations have yielded some intriguing results which may point to new physics or a need for a better understanding of galaxy formation. \cite{2022arXiv220712446L} have reported several galaxy candidates in $z \in [7,10] $ with stellar masses $\sim \mathcal{O} (10^{10}M_\odot)$. These galaxies may be in tension with the Lambda cold dark matter ($\Lambda$CDM) cosmology~\citep{Boylan-Kolchin:2022kae,Lovell:2022bhx,Haslbauer:2022vnq}. \cite{Haslbauer:2022vnq} show that the predictions of a few simulations~\citep{2015MNRAS.446..521S,2018MNRAS.473.4077P,2019ComAC...6....2N} do not match with these observations. However, these results have a few caveats: these redshifts are photometric, and only a few have been confirmed spectroscopically. Refs.~\cite{Bouwens:2022gqg,Ferrara:2022dqw,2022arXiv220802794N,2022arXiv221003754K,2022arXiv220801816Z,2022arXiv220814999E,2023arXiv230412347A,2023arXiv230413721A} discuss the possible uncertainties in photometric redshift measurements for these galaxy candidates at such extreme distances. There have also been attempts to explain this tension by considering beyond $\Lambda$CDM cosmologies. For example, an early dark energy component in the Universe can lead to an early structure formation; therefore, it may ease this tension~\citep{Klypin:2020tud,Boylan-Kolchin:2022kae}. Other works suggest that the presence of primordial black holes or axion miniclusters~\citep{Liu:2022bvr,Hutsi:2022fzw,Yuan:2023bvh,Dolgov:2023eqt}, fuzzy dark matter~\citep{Gong:2022qjx}, primordial non-Gaussianity~\citep{Biagetti:2022ode}, cosmic strings~\citep{Jiao:2023wcn}, and other new physics scenario~\citep{Lovyagin:2022kxl} as a possible solution to this tension. Additionally, \cite{Wang:2022jvx} study the modified gravity model, dynamical dark energy, and dark matter-baryon interaction and show that these models fail to resolve this discrepancy within the allowed model parameters from Planck cosmic microwave background (CMB) observations. Additionally, a better understanding of galaxy formation physics~\citep{2023arXiv230304827D,yung2023ultrahighredshift,Prada:2023dix,2023arXiv230413890C} or busty star formation~\citep{2023arXiv230502713S,2023arXiv230715305S} can also explain the data (but see~\cite{Pallottini:2023yqg}). Spectroscopic confirmation of these galaxies can validate the discrepancy between $\Lambda$CDM cosmology and JWST observations. A few works have also studied the implications of JWST observed galaxies on various dark matter models~\citep{Maio:2022lzg,Dayal:2023nwi}.

Besides photometric surveys, several works have reported a few spectroscopically confirmed galaxies using observations from JWST Advanced Deep Extragalactic Survey (JADES)~ \citep{2022arXiv221204568C,Robertson:2022gdk,2023arXiv230207256B}, CEERS survey~\citep{ ArrabalHaro2023,2023arXiv230405378A,Fujimoto:2023orx,Kocevski2023, Larson2023, Sanders2023,Tang2023},  and other early data releases~\citep{Isobe2023,Jung2022,Nakajima2023}.
JADES spectroscopic survey has reported $5$ galaxies with stellar masses $\gtrsim 10^8 M_\odot$ at $z>10$~\citep{2022arXiv221204568C,Robertson:2022gdk,2023arXiv230207256B}. These galaxies are: JADES-GS-z10-0 at $z = 10.38^{+0.07}_{-0.06}$; JADES-GS-z11-0 at $z = 11.58\pm 0.05$; JADES-GS-z12-0 at $z = 12.63^{+0.24}_{-0.08}$; JADES-GS-z13-0 at $z = 13.20^{+0.04}_{-0.07}$; and JADES-GN-z11 at $z = 10.6034\pm 0.0013$. \cite{2022arXiv221212804K} and \cite{2023arXiv230413755M} compared these galaxies to $\Lambda$CDM simulations and found that they are consistent.

JWST observations of high-redshift galaxies can have important implications for astrophysical and cosmological models. It is important to study this thoroughly to find new physics signals. Here, we explore the implications of JWST observations on the primordial power spectrum. We use a modified form of the primordial power spectrum with a blue/\,red tilt on small length scales. Modified power spectrum can arise in various beyond standard cosmological models. For instance, inflation models with running inflaton mass or modified dispersion relation can give rise to an enhancement in the primordial power spectrum at small scales~\citep{Covi:1998mb,Martin:2000xs,Hirano:2015wla}. Additionally, inflation models with an inflection point~\cite{Garcia-Bellido:2017mdw,Ballesteros:2017fsr,Germani:2017bcs} or bump~\cite{Mishra:2019pzq} in the inflaton potential also result in an enhanced power spectrum at small scales. It is also found that multi-field inflation models, where the inflaton field is coupled to a secondary field~\cite{Braglia:2020eai} or those features a mild waterfall phase in hybrid inflation~\cite{Gong:2010zf,Clesse:2015wea}) or rapid turns in field space~\cite{Palma:2020ejf} can also lead to an enhanced power spectrum. Using CEERS photometric observations, along with spectroscopic updates for some of their candidates, we compute the cumulative comoving stellar mass density (CCSMD) and show that a very high star formation efficiency ($\epsilon$) is required to explain these observations within $\Lambda$CDM. Such high star formation efficiencies are unlikely in various theoretical scenarios. We show that a blue-tilted power spectrum can potentially explain these observations, with a low to moderate star formation efficiency. In the second part of this work, we investigate the implications of JADES spectroscopic observations on the primordial power spectrum. For this purpose, we use the star formation efficiency derived from the previous Spitzer observations of galaxies with $z \approx 10$~\citep{2021ApJ...922...29S}. It is important to note that the estimated star formation efficiency is an order of magnitude smaller than that required to explain the measurements by \cite{2022arXiv220712446L}. JADES spectroscopic observations imply lower limits on the cumulative comoving galaxy number density (CCGND) at different redshifts. We use them to put the strongest constraints on the red tilt in the power spectrum over some length scales.

\section{Halo Mass Function}\label{sec:hmf}
The halo mass function (HMF) is defined as the number density ($n$) of DM haloes per unit mass:
\begin{equation}\label{eq:hmf}
	\frac{dn}{d \ln M} = M \frac{\rho_0}{M^2} f(\sigma) \left| \frac{d \ln \sigma}{d \ln M} \right| 	\,,
\end{equation}
where $\sigma$ and $\rho_0$ are the mass variance of smoothed linear matter density field in a sphere of radius $R$ and the mean density of the Universe, respectively. The radius $R$ is related to the mass M within the sphere as $M = \frac{4\pi\rho_0}{3}R^3$. The mass variance ($\sigma$) depends on the linear matter power spectrum, $P(k)$, and is given by
\begin{equation}
	\label{eq:sig}
	\sigma^2(R) = \frac{1}{2\pi^2}\int_0^\infty{k^2P(k)W^2(kR)dk}\,,
\end{equation}
where $k$ is the wavenumber and $W(kR)$ is a filter function in Fourier space; we use a top-hat filter function.  We use the fitting function, $f(\sigma)$, obtained using the Press-Schechter formalism~\citep{Press:1973iz} and including the corrections for ellipsoidal collapse~\citep{Sheth:1999mn}:
\begin{equation}
	f(\sigma) =A\sqrt{\frac{2a}{\pi}}\left[1+\left(\frac{\sigma^2}{a\delta_c^2}\right)^p\right]\frac{\delta_c}{\sigma}\exp\left[-\frac{a\delta_c^2}{2\sigma^2}\right]\,,
\end{equation}
where $\delta_c$ is the critical overdensity for collapse, $A=0.3222$, $a=0.707$, and  $p=0.3$. 

The linear matter power spectrum can be expressed as $P (k) = P_{\rm prim} (k) T^2 (k)\, ,$ where $P_{\rm prim} (k)$ is the primordial power spectrum and $T(k)$ is the transfer function, which governs the evolution of sub-horizon modes. In standard cosmology, $P_{\rm prim} (k) \propto k^{n_s}$, where $n_s$ is the spectral index. In this work, we study a modified primordial power spectrum where it deviates from the standard primordial power spectrum at small length scales with a model agnostic form:
\begin{align}
	P_{\rm prim} (k) &\propto k^{n_s}, ~~~~~ {\rm for}\, k < k_p, \\
	& \propto  k_p^{n_s - m_s} k^{m_s},~~~~~ {\rm for}\, k>k_p\,.
\end{align}
The deviation from standard primordial power spectrum depends on the pivot scale, $k_p$, and the tilt ($m_s$) on scales $k>k_p$. For $m_s > n_s$, the power spectrum will be blue tilted on scales $k>k_p$, and it is red tilted if $m_s < n_s$. Alternatively, we can also model $P_{\rm prim} (k)$ with a non-zero running of spectral index~\citep{2020A&A...641A..10P}.

  \begin{figure*}
 	\begin{center}
 		\hspace{-0.5cm}
 		\includegraphics[width=16cm,height=6.5cm]{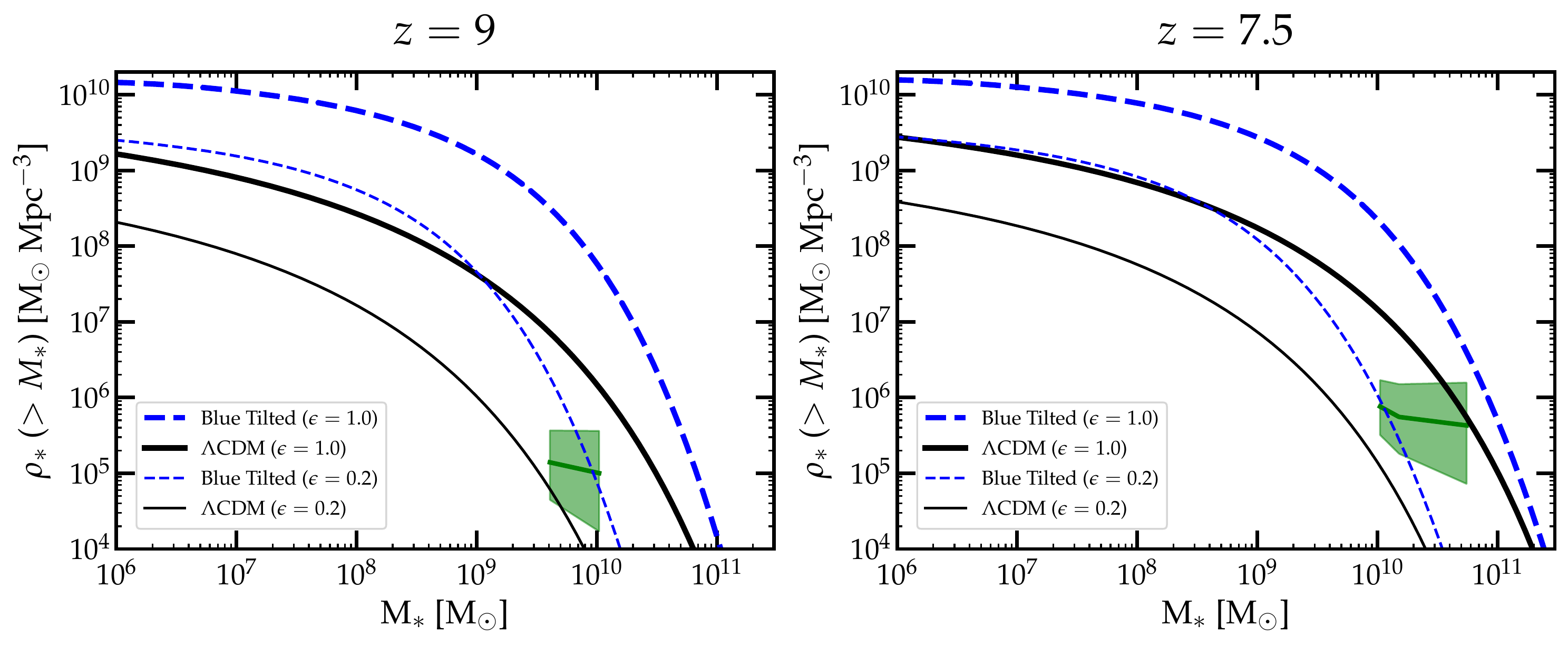}
 		\caption{ The CCSMD of galaxies with stellar mass content more than $M_{\ast}$ for $z=9$ ({\bf left-hand panel}) and $7.5$ ({\bf right-hand panel}). The black and blue curves are for the standard $\Lambda$CDM cosmology and the cosmology with a blue tilted primordial power spectrum ($k_p = 1~{\rm h\, Mpc}^{-1}$ and $m_s = 2.0$), respectively. The thick and thin curves are for $\epsilon =1 .0$ and $0.2$, respectively. The green bands represent the CCSMD that we have computed using observations by~\protect \cite{2022arXiv220712446L} and corresponding spectroscopic updates.}\label{fig:rho_z}
 	\end{center}
 \end{figure*} 
 \begin{figure*}
	\begin{center}
		\hspace{-0.5cm}
		\includegraphics[width=16cm,height=6.5cm]{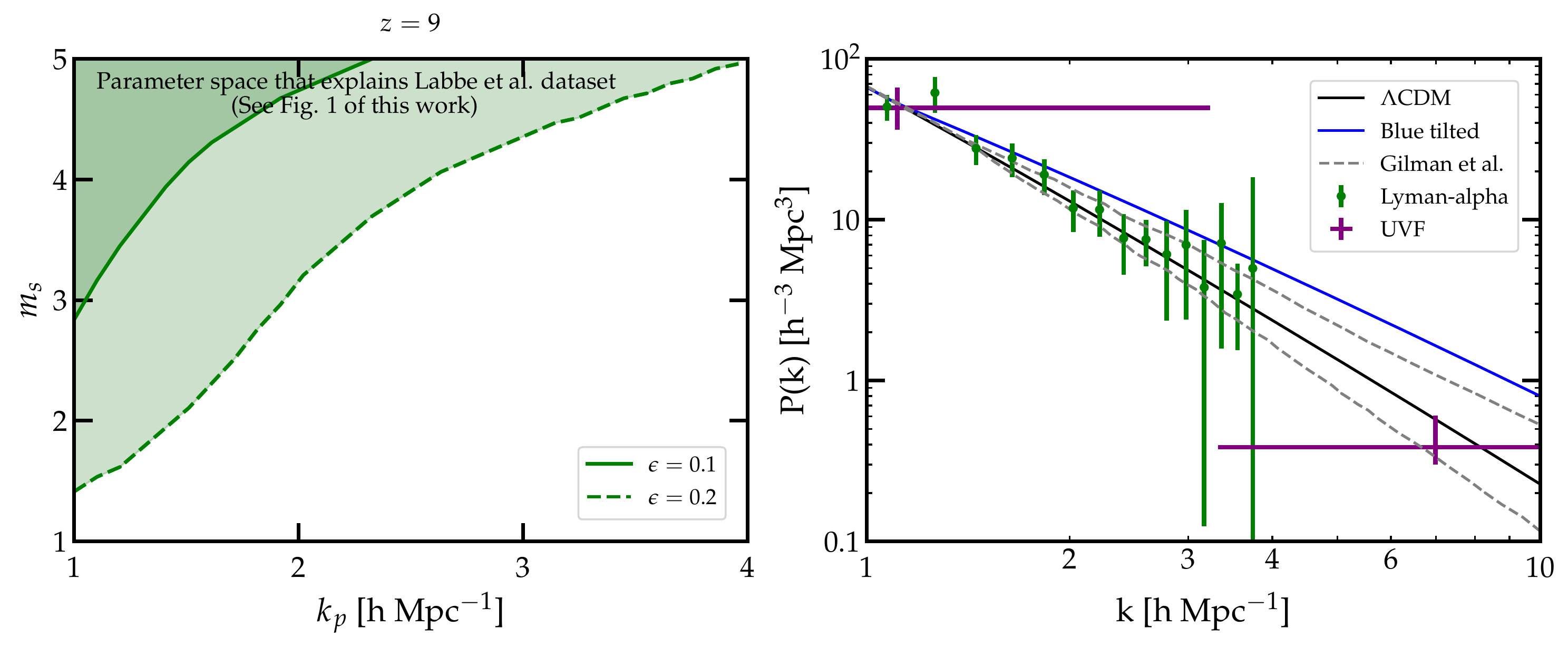}
		\caption{ {\bf Left-hand panel:} The shaded region represents the $k_p-m_s$ parameter space which predicts CCSMD consistent with JWST inferred $\rho_\ast (> M_{\ast})$ assuming $M_\ast$ fixed at the mass of the most massive galaxy candidate at $z \approx 9$ by~\protect\cite{2022arXiv220712446L}. The solid and dashed curves are for $\epsilon=0.1$ and $0.2$, respectively.
			{\bf Right-hand panel:} Matter power spectra computed at $z=0$ within the standard $\Lambda$CDM cosmology and cosmology with a blue tilted primordial power spectrum ($k_p = 1.1 \, {\rm h~ Mpc}^{-1}$ and $m_s = 1.53$) are shown by the black and blue solid lines, respectively, along with constraints from other measurements by \protect\cite{Viel:2004np,2019JCAP...07..017C,2019MNRAS.489.2247C, 2022ApJ...928L..20S,Gilman:2021gkj} .}\label{fig:b_const}
	\end{center}
\end{figure*} 

\section{Analysis and Results}\label{sec:res}
Following \cite{Boylan-Kolchin:2022kae}, we compute the CCSMD and CCGND assuming a modified form of the primordial power spectrum. We first calculate the cumulative comoving number density of haloes with masses above some threshold $M_{\rm halo}$ as
\begin{equation}\label{eq:n}
	n(> M_{\rm halo}, z) = \int_{M_{\rm halo}}^{\infty} dM \frac{dn (M,z)}{dM}\, ,
\end{equation}	
and the corresponding cumulative comoving mass density of haloes
\begin{equation}\label{eq:rho}
	\rho (> M_{\rm halo}, z) = \int_{M_{\rm halo}}^{\infty} dM M \frac{dn (M,z)}{dM}\,.
\end{equation}	
These relations can be directly translated to compute the CCGND, $n_\ast(> M_{\ast}, z)$, and CCSMD, $\rho_\ast (> M_{\ast}, z)$, with stellar masses greater than $M_{\ast}$, assuming $M_\ast = \epsilon f_b M_{\rm halo}$, where $f_b = \Omega_b/ \Omega_m$\footnote{We have fixed the cosmological parameters at their best-fit value obtained from the Planck CMB observations throughout this work~\citep{Planck:2018vyg}. The values are as follows: matter density fraction $\Omega_m = 0.3153$,  baryon density fraction $\Omega_b = 0.0493$, Hubble constant $H_0 = 67.36$ km s$^{-1}$ Mpc$^{-1}$, $n_s=0.9649$, and $\sigma_8=0.8111$.}. The exact $\epsilon$ value depends on the star formation physics; however, it satisfies the inequality $\epsilon \leq 1$ and we assumes it to be constant over a redshift range. The CCSMD is given as $\rho_\ast (> M_{\ast}, z)  = \epsilon f_b \,\rho (> M_{\rm halo}, z)$.

{\it Photometric candidates: }Recently, \cite{2022arXiv220712446L} have reported 13 galaxy candidates, which were identified in the JWST CEERS program, with stellar masses $\sim 10^9 - 10^{11} M_\odot$ in the photometric redshift range $z \sim 6.5 - 9.1$. Some of the galaxies have also been confirmed spectroscopically~\citep{2023arXiv230405378A,Fujimoto:2023orx}. We work with updated mass and redshift measurements for these galaxies. We have also accordingly updated the CCSMD computed by \cite{2022arXiv220712446L} in two redshift bins $z \in [7,8.5]$  and $z \in [8.5,10]$, using the three most massive galaxies, which are shown in Fig.~\ref{fig:rho_z} by green bands. The uncertainties in CCSMD includes both Poisson errors and cosmic variance. Cosmic variance is computed using a web calculator~\citep{2008ApJ...676..767T} and is approximately 30$\%$. In the two bins, the most massive galaxies are at $z \approx 7.5 $ and $9$ with $M_\ast \approx 10^{11} M_\odot$ and $10^{10} M_\odot$, respectively. We compute CCSMD within $\Lambda$CDM for $\epsilon = 1.0$ and $0.2$ at redshifts $z=7.5$ and $9$, shown by black lines in Fig.~\ref{fig:rho_z}. We also compute CCSMD assuming a blue-tilted primordial power spectrum with $k_p = 1\,{\rm h\, Mpc}^{-1}$ and $m_s = 2.0$, displaying them as a function of $M_{\ast}$ by blue dashed curves in Fig.~\ref{fig:rho_z}. We have used {\tt HMF} code in our work~\citep{Murray:2013qza,2014ascl.soft12006M}.

It is evident from Fig.~\ref{fig:rho_z}, that CCSMD within $\Lambda$CDM with $\epsilon = 0.2$ is not consistent with the JWST inferred CCSMD. More specifically, the JWST inferred CCSMD at $z \approx 9$ and $z \approx 7.5$ is consistent with the predictions within the standard cosmology for $\epsilon \gtrsim 0.45$ and $\epsilon \gtrsim 0.95$. When we consider the 1$\sigma$ uncertainties, slightly smaller values of $\epsilon$ will be consistent with the JWST inferred CCSMD. However, for both cases, the required $\epsilon$ is either inconsistent or marginally consistent with the plausible theoretical values ($\epsilon \lesssim 0.32$)~\citep{Gribel:2017rpq,Tacchella:2018qny,Behroozi:2020jhj}, pointing towards a tension between the JWST observations and standard cosmology (see also \cite{Boylan-Kolchin:2022kae}). We can see from Fig.~\ref{fig:rho_z}, that a smaller $\epsilon$ is required to explain the JWST inferred CCSMD in cosmology assuming a blue tilted primordial power spectrum than that required within the standard cosmology. 
Therefore, this tension can be eased by a blue-tilted primordial power spectrum. From Fig.~\ref{fig:rho_z}, the CCSMD obtained for a blue tilted primordial power spectrum with $k_p = 1\, {\rm h~ Mpc}^{-1}$ and $m_s = 2.0$ can be consistent with CCSMD inferred from JWST, at $z=9$ with $\epsilon =0.2$. For $z=7.5$, a slightly larger $\epsilon$ will be required.
A larger $m_s$ can lead to an even large value of $\rho_\ast (> M_{\ast})$, but we cannot choose any values of $m_s$ and $k_p$ as these will be limited by the constraints on matter power spectrum from various measurements \citep{Viel:2004np,2019JCAP...07..017C,2019MNRAS.489.2247C, 2022ApJ...928L..20S,Gilman:2021gkj}. 

\begin{figure*}
	\begin{center}
		\hspace{-1cm}
		\includegraphics[width=16cm,height=6.5cm]{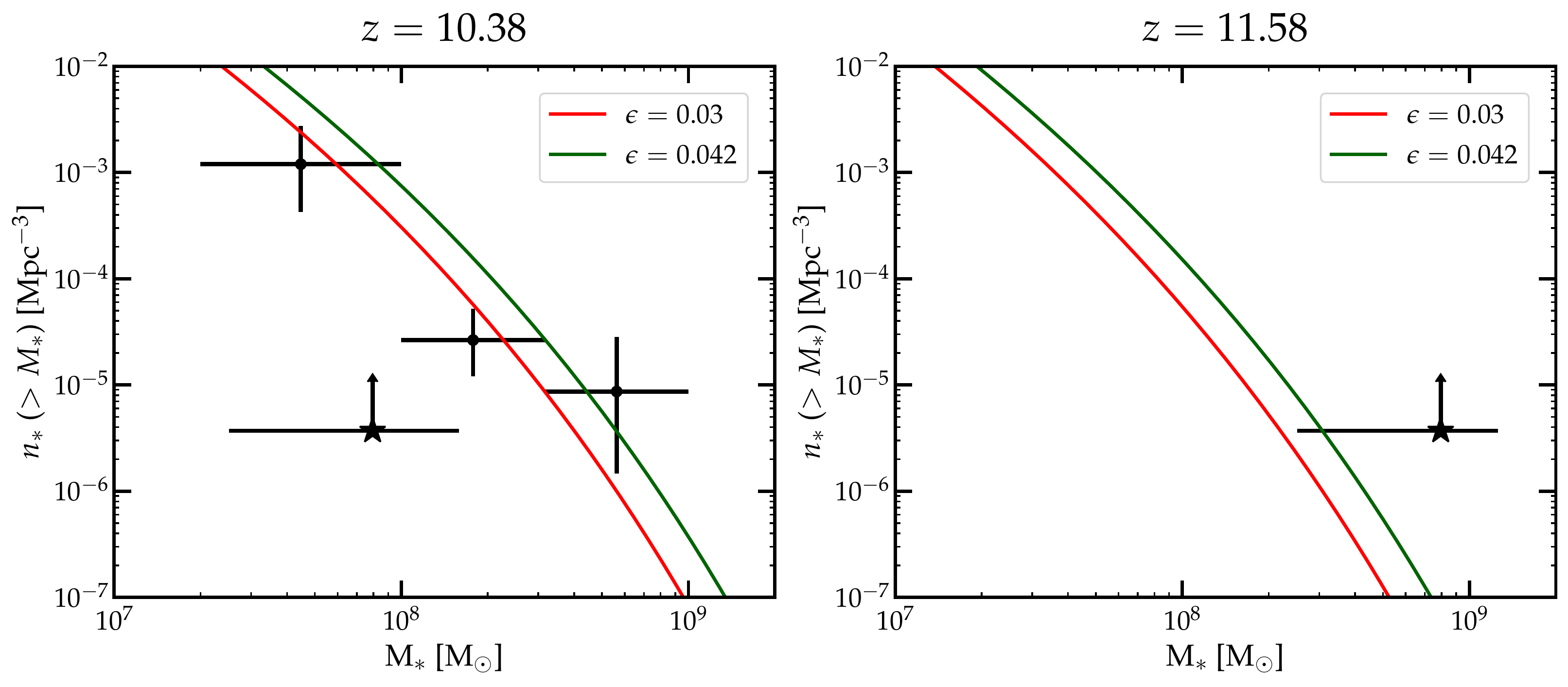}
		\caption{ The CCGND of galaxies with stellar masses more than $M_{\ast}$ for $z=10.38$ ({\bf left-hand panel}) and $11.58$ ({\bf right-hand panel}) within the $\Lambda$CDM cosmology. The red and green curves are obtained with $\epsilon =0.03$ and $0.042$, respectively, in order to match the data obtained by~\protect\cite{2021ApJ...922...29S}, which are shown by black data points. The black stars represent the CCGND inferred from the recent JWST observations~\citep{2022arXiv221204568C,Robertson:2022gdk,2022arXiv221212804K}.}\label{fig:n_var}
	\end{center}
\end{figure*} 

\begin{figure}
	\begin{center}
		\hspace{-0.75cm}
		\includegraphics[width=7cm,height=6.5cm]{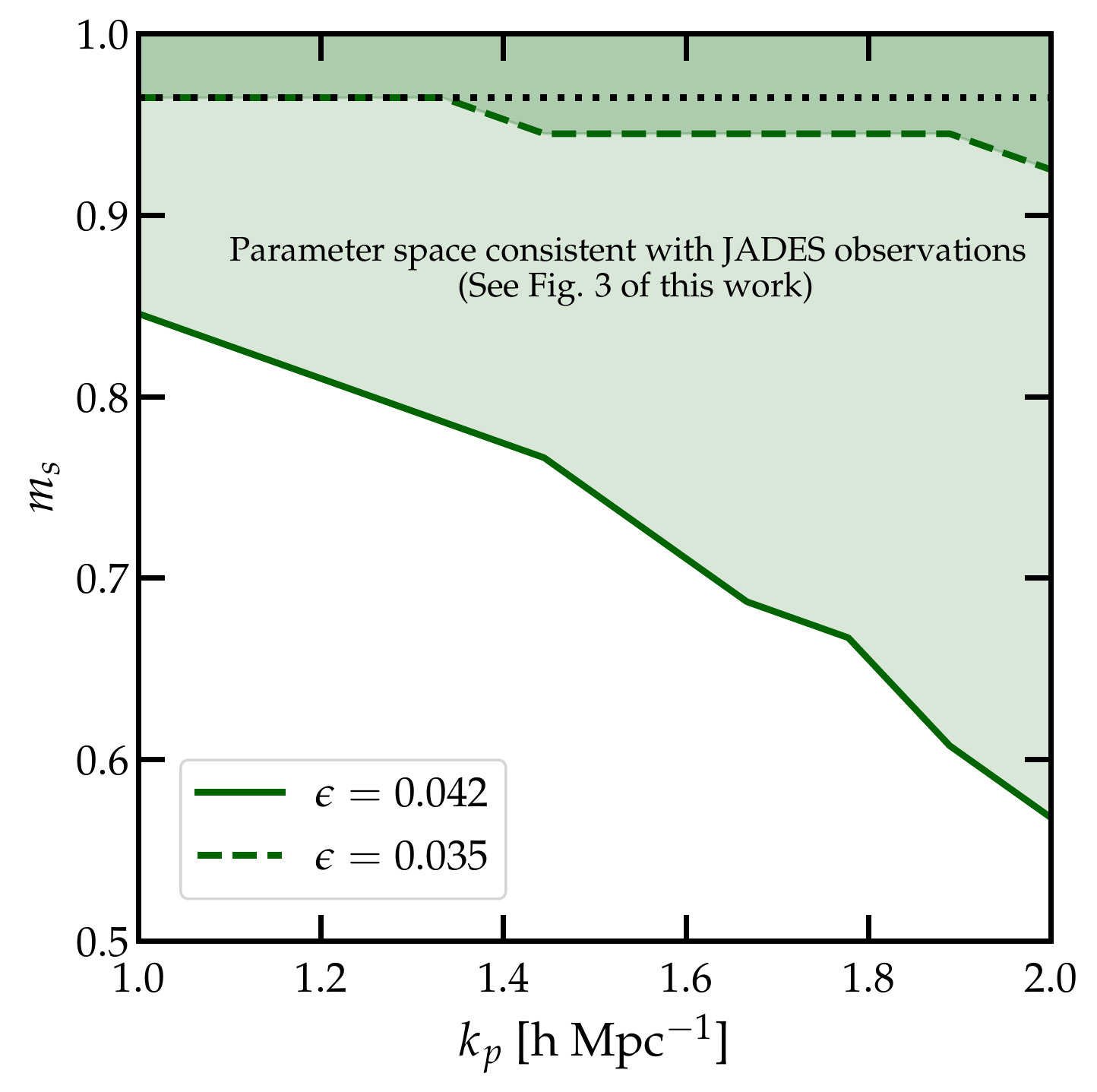}
	\end{center}
	\caption{The shaded regions show the $k_p-m_s$ parameter space consistent with JADES observations. For all the values in the shaded region, the predicted CCGND for a red-tilted power spectrum is consistent with that inferred for the galaxies observed in the JADES survey. Parameter space below the green shaded regions are ruled out by JADES observations. The solid and dashed curves are for $\epsilon=0.042$ and $0.035$, respectively. The black dotted line shows the spectral index $n_s =0.9649$, within $\Lambda$CDM cosmology.}\label{fig:const_red}
\end{figure}

We perform a scan of $m_s$ and $k_p$ values to find the parameter space that can achieve the CCSMD inferred from the JWST observations. We vary $k_p \in [1,5]$ h Mpc$^{-1}$ and $m_s \in [0.9649, 5]$, with $\epsilon = 0.1 $ or $0.2$, and compute the corresponding CCSMD ($\rho_\ast (> M_{\ast})$) with $M_\ast$ fixed at the maximum value of JWST inferred stellar mass at $z\approx9$. To infer the parameter space consistent with CEERS observations, we compare the computed CCSMD with that inferred from CEERS observation. If the computed CCSMD is greater or equal to the lowest (at 68$\%$ c.l.) value of the CCSMD inferred from CEERS observations, we select the $k_p$ and $m_s$ values; otherwise, we reject them. The selected $k_p$ and $m_s$ values represent the parameter space consistent with CEERS observations and are shown by the green shaded regions in the left panel of Fig.~\ref{fig:b_const}. The solid and dashed lines correspond to $\epsilon =0.1$ and $0.2$, respectively. It is evident from the figure that a larger tilt is required to explain JWST inferred $\rho_\ast (> M_\ast)$ for $\epsilon=0.1$ than that required for $\epsilon=0.2$. 
The shaded regions may be in conflict with the 68$\%$ c.l. error bars presented in \cite{2022ApJ...928L..20S,Gilman:2021gkj}. As an example, the linear matter power spectrum obtained assuming $\Lambda$CDM transfer function and a blue tilted primordial power spectrum with $k_p = 1.1 \, {\rm h~ Mpc}^{-1}$ and $m_s = 1.53$, which can explain the \cite{2022arXiv220712446L} result, is shown in the right panel of Fig.~\ref{fig:b_const}. There are a few ways to address this issue: 1. JWST results are slightly revised, although they still remain in tension with $\Lambda$CDM predictions. 2. Our inferred $m_s - k_p$ region is probably consistent with the 95$\%$ c.l. regions of other measurements. 3. Even if the inferred $m_s - k_p$ region for $\epsilon =0.2$ is inconsistent with the other measurements, some other $m_s - k_p$ values with a slightly large $\epsilon$ will still be allowed. Therefore, a blue-tilted spectrum will still ease the tension though it may not resolve it completely. 4. Some new physics changes the transfer function such that only a small blue tilt is required. Finally, it is also possible that star formation efficiency is higher for massive galaxies at high redshifts~\citep{2023arXiv230304827D}. The required blue tilt may also impact the reionization epoch~\citep{Gong:2022qjx}, which future observations can probe. In addition, we would also like that a blue-tilted power spectrum will also exacerbate some of the small-scale tensions in $\Lambda$CDM cosmology. A blue-tilted power spectrum will imply a larger number of satellite galaxies as compared to that predicted in $\Lambda$CDM cosmology, which will worsen the missing satellites problem. However, since the current JWST observations do not  confirm the existence of a blue-tilted power spectrum at small scales, we do not address this problem in this work.

\begin{figure*}
	\begin{center}
		\includegraphics[width=16cm,height=6.5cm]{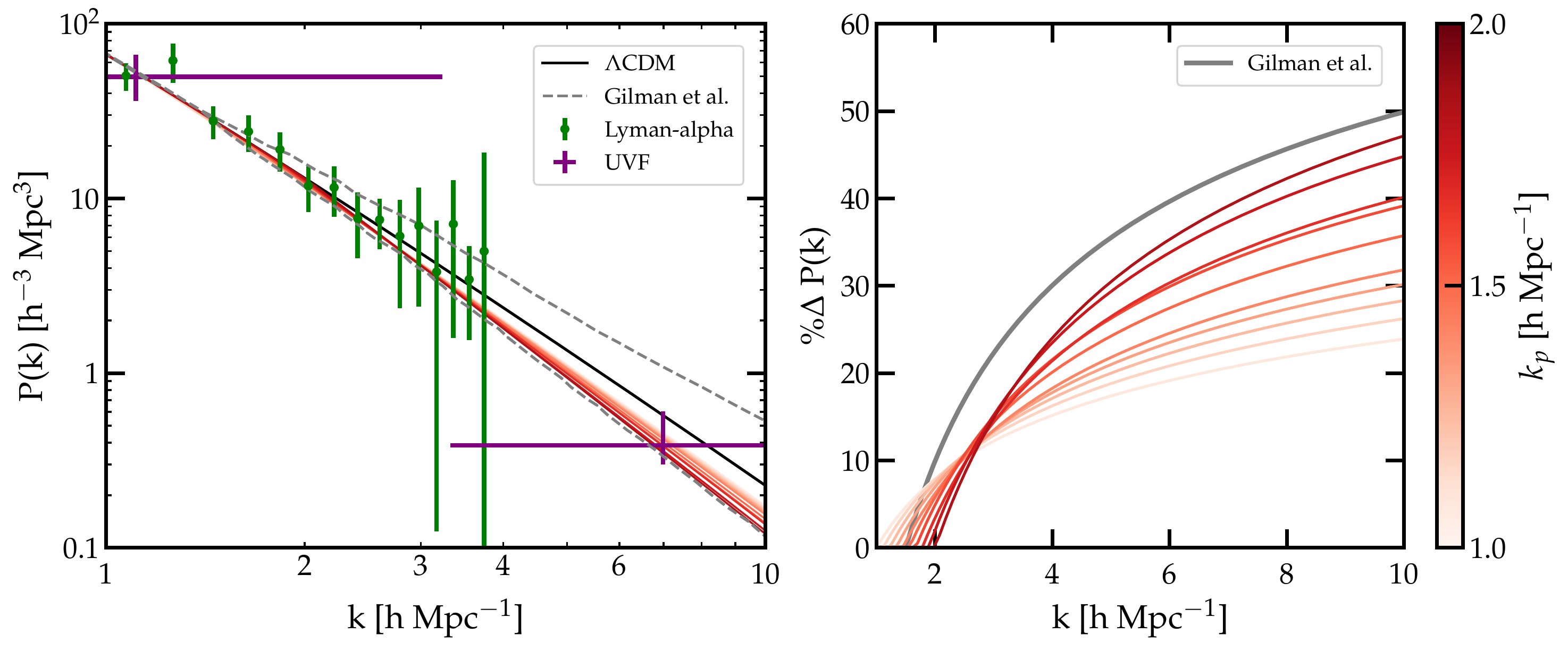}
		\caption{ {\bf Left-hand panel:} Matter power spectra computed at $z=0$ within the standard $\Lambda$CDM cosmology (black solid line) and cosmology with a red tilted primordial power spectrum (red solid lines). The multiple red lines are obtained for $k_p \in [1,2]$ h~Mpc$^{-1}$ and show the lower limits on the red-tilt for each $k_p$ from JWST observations with $\epsilon=0.042$. 
			{\bf Right-hand panel:} The maximum allowed percentage change in $P(k)$ with a red tilted primordial power spectrum as compared to that in $\Lambda$CDM cosmology, $\%\Delta P(k)$,
			is shown by multiple red curves for $k_p \in [1,2]$ h~Mpc$^{-1}$ and corresponding allowed $m_s$ values. The gray curve represents the maximum percentage deviation allowed in $P_{\Lambda \rm CDM} (k)$ by the constraints obtained in~\protect\cite{Gilman:2021gkj}. These plots show that JWST observations impose the strongest constraints on scales $k \sim 2-7$ h~Mpc$^{-1}$. }\label{fig:pk-var}
	\end{center}
\end{figure*}

{\it Spectroscopically confirmed galaxies: } Recently \cite{2022arXiv221204568C,Robertson:2022gdk} have reported $4$ galaxies from the JADES survey with spectroscopically confirmed redshifts. \cite{2022arXiv221212804K} report the lower limit on CCGND inferred from these observations, shown in Fig.~\ref{fig:n_var} by black stars. There are also other estimates of CCGND at $z\sim 10$ obtained using Spitzer data by \cite{2021ApJ...922...29S}, which are shown by black points with error bars in Fig.~\ref{fig:n_var}. There are few other spectroscopic measurements, but our work only uses these observations. For a given power spectrum, the CCGND depends on $\epsilon$. We first compute the CCGND at $z\sim 10$, within the standard $\Lambda$CDM cosmology, and compare it with the results obtained by \cite{2021ApJ...922...29S} to find a range of $\epsilon \in [0.03, 0.042]$. The CCGND as a function of stellar mass for these two $\epsilon$ values are shown by the green and red curves in Fig.~\ref{fig:n_var}. It is evident from Fig.~\ref{fig:n_var} that the CCGND within the standard cosmology and $\epsilon$ in this range will be consistent with the results of \cite{Robertson:2022gdk} and \cite{2021ApJ...922...29S}. Since a red-tilted primordial power spectrum predicts a smaller CCGND, we can use this observational data to constrain the red-tilt for a given $\epsilon$. As we have already obtained a range of $\epsilon$ values which gives results consistent with the standard cosmology, we fix $\epsilon \approx 0.03$ and $0.042$ for these galaxies. Note that these $\epsilon$ values are an order of magnitude smaller than that required to explain the \cite{2022arXiv220712446L} results. These differences may arise due to environmental effects; more observations are required to clarify the situation. Next, we compare the CCGND obtained for a red-tilted power spectrum at $z=11.38, 12.53$, and $13.32$ with those inferred from the JWST observations by \cite{Robertson:2022gdk} at these redshifts. The parameter space of $m_s$ and $k_p$ that predicts a smaller CCGND than that inferred from the JWST observations at these redshifts will be disallowed. We perform a similar parameter scan analysis with JADES observations as done with CEERS observations to find the constraints on the red-tilted power spectrum.  We vary $k_p \in [1,2]$ h Mpc$^{-1}$ and $m_s \in [0.9649, 0]$, with $\epsilon = 0.035 $ or $0.042$, and compute the corresponding CCGND ($n_\ast (> M_{\ast})$), with $M_\ast$ fixed at the minimum value of JWST inferred stellar mass of the galaxies.  Next, we compare the computed CCGND with that inferred from JADES observations to obtain the parameter space constrained by JADES observations.  Since JADES observation provides a lower limit on CCGND, the $k_p$ and $m_s$ values for which the computed CCGND is smaller than JADES inferred CCGND will not be allowed.  Hence, if the computed CCGND is smaller than the JADES inferred CCGND, we reject the corresponding values to obtain the constrained $k_p-m_s$ parameter space.  The galaxy at $z=11.38$ puts the most stringent constraint; the disallowed region is shown in Fig.~\ref{fig:const_red} by the region below green shades.  The curves with solid and dashed lines correspond to $\epsilon=0.035$ and $0.042$, respectively. It is evident from Fig.~\ref{fig:const_red} that only a tiny red tilt in the power spectrum is allowed from JWST observations. The constraints will be more stringent or relaxed with a smaller or larger value of $\epsilon$. Since the exact $\epsilon$ value depends on complex astrophysical processes, and there is a degeneracy between $\epsilon$ and the power spectrum parameters, we can only constrain the power spectrum parameters for a given value of $\epsilon$.

We also study the implications of these constraints on the matter power spectrum, assuming the $\Lambda$CDM transfer function. If the transfer function changes due to the presence of new physics, the constraints on the matter power spectrum will also change accordingly. We plot the linear matter power spectra with the maximum allowed red tilts for $k_p \in [1,2]$ and $\epsilon =0.042$ in the left panel of Fig.~\ref{fig:pk-var} by multiple red curves. We also show the constraints from \cite{Viel:2004np,2019JCAP...07..017C,2019MNRAS.489.2247C,2022ApJ...928L..20S,Gilman:2021gkj}; all of them are consistent with $\Lambda$CDM cosmology. 
It can be seen from Fig.~\ref{fig:pk-var} that the matter power spectra obtained with the maximum allowed red tilts for $k_p \in [1,2]$ and $\epsilon =0.042$ lie within the 68$\%$ c.l. band reported in \cite{Gilman:2021gkj}. Recently, \cite{Esteban:2023xpk} have also constrain the matter power spectrum at small scales using Milky Way satellite velocities.  We also plot the percentage deviation of matter power spectrum obtained with a red-tilt from that predicted in standard cosmology, defined as $\%\Delta P(k) = \frac{P_{\Lambda \rm CDM} (k) - P_{\rm rt}(k)}{P_{\Lambda \rm CDM} (k)} \times 100$,
where $P_{\Lambda \rm CDM} (k)$ and $P_{\rm rt}(k)$ are matter power spectra within $\Lambda$CDM cosmology, and cosmology with red-tilted primordial power spectrum, respectively. The $\%\Delta P(k)$ is shown in the right panel of Fig.~\ref{fig:pk-var} by red curves. It is clear from this figure that $\%\Delta P(k)$ is less than the maximum deviation allowed in $P(k)$ by the constraints obtained in \cite{Gilman:2021gkj}, shown by a gray line. From Fig.~\ref{fig:pk-var}, we find that observations of high redshift massive galaxies can give the strongest constraints on matter power spectrum on scales $k \sim 2-7$ h~Mpc$^{-1}$. If future JWST observations reveal a higher number density of galaxies at these masses and redshifts, then this will imply an even stronger constraint on the matter power spectrum; thus demonstrating that such observations may be the best probe of the matter power spectrum at $1 \lesssim k \lesssim 10$ h~Mpc$^{-1}$.

\section{Conclusions}\label{sec:con}
JWST has already provided very interesting, and perhaps surprising, results by observing several massive galaxy candidates at high redshifts. Many of these candidates are detected photometrically, and some of them have been confirmed as galaxies using spectroscopy. 

Detections of some galaxy candidates in the CEERS survey with stellar masses $\gtrsim 10^{10} M_\odot$ at photometric redshifts $z\gtrsim7$ have caught much attention. If all of them are confirmed spectroscopically, these can seriously challenge the $\Lambda$CDM cosmology: how do such massive galaxies form so early in the timeline of the Universe? A solution to this challenge may come from a better understanding of galaxy formation physics or beyond $\Lambda$CDM cosmology. This work shows that a blue-tilted primordial power spectrum can ease this tension. Recently, \cite{Hirano:2023auh} did simulations with a blue-tilted power spectrum and found their results consistent with our results. However, $k_p -m_s$ parameter space that eases this tension is highly constrained from various astrophysical observations. Soon after this work, \cite{Sabti:2023xwo} performed an analysis by assuming a Gaussian enhancement in the power spectrum and found that the enhancement required to explain \cite{2022arXiv220712446L} observations will conflict with previous constraints on these scales by \cite{2022ApJ...928L..20S} (see Fig. \ref{fig:b_const} of this work). Additionally, there are large systematic and statistical uncertainties in these measurements. If future JWST data can shrink these uncertainties and the redshifts are spectroscopically confirmed, these observations will be an important tool in probing the power spectrum at small length scales, and will be a discovery tool for new physics.

We also study the impact of spectroscopically confirmed galaxies observed in the JADES survey by JWST on the primordial power spectrum. These galaxies are consistent with the $\Lambda$CDM cosmology. We use four spectroscopically confirmed galaxies at $z>10$ to constrain a red-tilted primordial power spectrum. We find that the most stringent constraint on the red tilt of the primordial power spectrum comes from the observation of a galaxy at redshift $z=11.38$. These constraints depend on $\epsilon$: for smaller values of $\epsilon$, the constraints will be stringent, whereas they will weaken if $\epsilon$ is larger. Using the $\epsilon$ values derived from Spitzer data, we find the most stringent constraint on the matter power spectrum at $k \sim 2-7$ h~Mpc$^{-1}$. In the future, more spectroscopically confirmed galaxies at such high redshifts by JWST can further strengthen these constraints.

We would like to point out that the JADES measurements only provide a lower limit on the CCGND, hence, it will be formally consistent with a blue-tilted power spectrum if we ignore other measurements that constrain the power spectrum at these scales. However, it is not appropriate to exclusively compare only the CEERS and JADES observations while neglecting the other measurements at these length scales. When CEERS and JADES observations are combined with earlier measurements, it is apparent that there is a discrepancy between the two datasets: CEERS is probably inconsistent with $\Lambda$CDM cosmology, whereas JADES is consistent with it. Consequently, it is impossible to find a modified power spectrum that explains both sets of observations while also remaining consistent with previous measurements. As these datasets have inconsistency, we do not employ them to {\it simultaneously} constrain the power spectrum. Our aim is to highlight the significance of JWST observations as a potential power spectrum probe. Therefore, we use these observations to study their implications on the power spectrum independently.

 Our work shows that near-future observations of massive galaxies at high redshifts by JWST can either teach us more about galaxy formation physics and star formation efficiency in the early Universe or discover new physics beyond $\Lambda$CDM physics.

%%%%%%%%%%%%%%%%%%%%%%%%%%%%%%%%%%%%%%%%%%

\section*{Acknowledgements}
 PP acknowledges the IOE-IISc fellowship program for financial assistance. RL acknowledges financial support from the Infosys foundation (Bangalore), institute start-up funds, and the Department of Science and Technology (Govt. of India) for the grant SRG/2022/001125. The authors also acknowledge valuable discussions with Tom Abel, Susmita Adhikari, Pablo Arrabal Haro, Arka Banerjee, Michael Boylan-Kolchin, Mousumi Das, Benjamin Keller, Ivo Labb\'e, Subhendra Mohanty, and Prateek Sharma. We thank Arka Banerjee, Michael Boylan-Kolchin, Mousumi Das, Benjamin Keller, Subhendra Mohanty, Nirupam Roy, and Prateek Sharma for comments on the manuscript. 

%%%%%%%%%%%%%%%%%%%%%%%%%%%%%%%%%%%%%%%%%%%%%%%%%%
\section*{Data Availability}
There are no new data associated with this article.
 


\bibliographystyle{mnras}
\bibliography{jwst2} % if your bibtex file is called example.bib

%\begin{thebibliography}{99}
%\end{thebibliography}

%%%%%%%%%%%%%%%%%%%%%%%%%%%%%%%%%%%%%%%%%%%%%%%%%%

%%%%%%%%%%%%%%%%% APPENDICES %%%%%%%%%%%%%%%%%%%%%

%\appendix

%%%%%%%%%%%%%%%%%%%%%%%%%%%%%%%%%%%%%%%%%%%%%%%%%%

% Don't change these lines
\bsp	% typesetting comment
\label{lastpage}
\end{document}